\documentclass[aps,prb,final,10pt,twocolumn,superscriptaddress,nofootinbib,citeautoscript,flushbottom]{revtex4-1}%
\usepackage{amssymb,amsmath,amsfonts}
\usepackage{graphicx,color}
\usepackage[squaren]{SIunits}
\usepackage[english]{babel}
\usepackage{tabularx}
\usepackage[]{units}

\newcolumntype{Y}{>{\centering\arraybackslash}X}

\newcommand{\dif}{\mathrm{d}}%
\newcommand{\abs}[1]{\lvert#1\rvert}%
\newcommand{\norm}[1]{\lVert#1\rVert}%

\begin{document}

\title{Helical paths, gravitaxis, and separation phenomena for mass-anisotropic self-propelling colloids: experiment versus theory}

\author{Andrew I. Campbell}
\affiliation{Department of Chemical and Biological Engineering, University of Sheffield, Sheffield S1 3JD, United Kingdom}

\author{Raphael Wittkowski}
\affiliation{Institut f\"{u}r Theoretische Physik, Westf\"alische Wilhelms-Universit\"{a}t M\"{u}nster, D-48149 M\"{u}nster, Germany}
\affiliation{Center for Nonlinear Science (CeNoS), Westf\"alische Wilhelms-Universit\"{a}t M\"{u}nster, D-48149 M\"{u}nster, Germany}

\author{Borge ten Hagen}
\affiliation{Physics of Fluids Group and Max Planck Center Twente, Department of Science and Technology, MESA+ Institute, and J. M. Burgers Centre for Fluid Dynamics, University of Twente, 7500 AE Enschede, The Netherlands}

\author{Hartmut L\"{o}wen}
\affiliation{Institut f\"{u}r Theoretische Physik II: Weiche Materie, Heinrich-Heine-Universit\"{a}t D\"{u}sseldorf, D-40225 D\"{u}sseldorf, Germany}

\author{Stephen J. Ebbens}
\email{Electronic address: s.ebbens@sheffield.ac.uk}
\affiliation{Department of Chemical and Biological Engineering, University of Sheffield, Sheffield S1 3JD, United Kingdom}

\date{\today}

\begin{abstract}
The self-propulsion mechanism of active colloidal particles often generates not only translational but also rotational motion. For particles with an anisotropic mass density under gravity, the motion is usually influenced by a downwards oriented force and an aligning torque. Here we study the trajectories of self-propelled bottom-heavy Janus particles in three spatial dimensions both in experiments and by theory. For a sufficiently large mass anisotropy, the particles typically move along helical trajectories whose axis is oriented either parallel or antiparallel to the direction of gravity (i.e., they show gravitaxis). In contrast, if the mass anisotropy is small and rotational diffusion is dominant, gravitational alignment of the trajectories is not possible. Furthermore, the trajectories depend on the angular self-propulsion velocity of the particles. If this component of the active motion is strong and rotates the direction of translational self-propulsion of the particles, their trajectories have many loops, whereas elongated swimming paths occur if the angular self-propulsion is weak. We show that the observed gravitational alignment mechanism and the dependence of the trajectory shape on the angular self-propulsion can be used to separate active colloidal particles with respect to their mass anisotropy and angular self-propulsion, respectively.
\end{abstract}

\maketitle

\section{Introduction}
\begin{figure}[htb]
\centering
\includegraphics[width=0.9\linewidth]{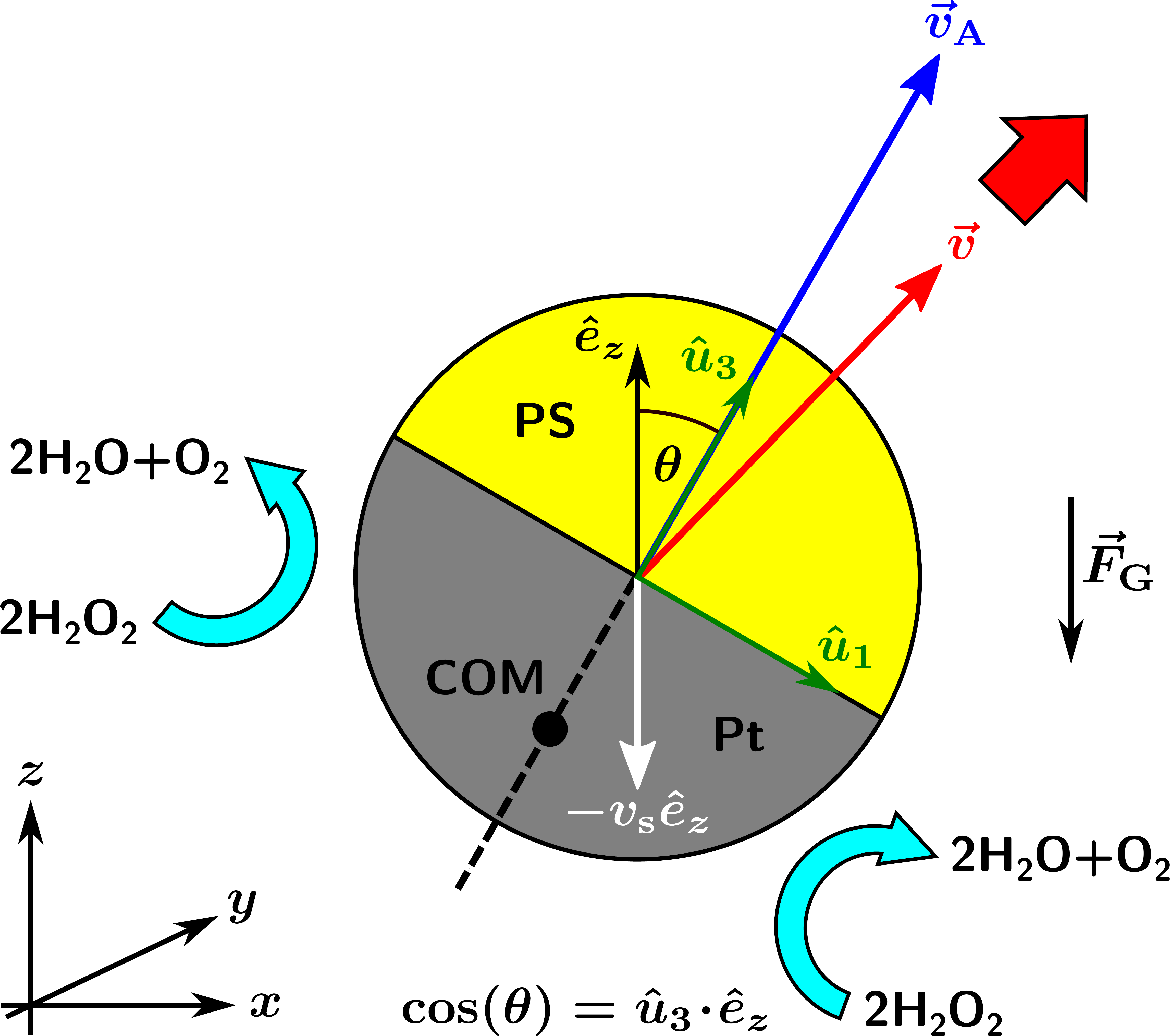}%
\caption{\label{fig:ParticleSchematic}Schematic of a self-propelled gravitactic Janus sphere in a hydrogen peroxide ($\mathrm{H}_{2}\mathrm{O}_{2}$) solution as used in our experiments. The platinum (Pt) cap of the polystyrene (PS) sphere causes a catalytic decomposition of $\mathrm{H}_{2}\mathrm{O}_{2}$ into water ($\mathrm{H}_{2}\mathrm{O}$) and oxygen ($\mathrm{O}_{2}$) leading to a self-propulsion of the particle with velocity $\vec{v}_{\mathrm{A}}$.\cite{Ebbens2011}    
If the gravitational force $\vec{F}_{\mathrm{G}}$ acts on the particle, its translational velocity $\vec{v}=\vec{v}_{\mathrm{A}}-v_{\mathrm{s}}\hat{e}_z$ deviates from the self-propulsion velocity $\vec{v}_{\mathrm{A}}$, where $v_{\mathrm{s}}$ is the sedimentation speed of the particle and $\hat{e}_{z}=(0,0,1)^\mathrm{T}$ is the unit vector parallel to the $z$ axis.
Because of the displaced center of mass (COM) the particle orientation is biased towards cap-down orientations, i.e., small tilt angles $\theta\in[0,\pi]$, and the particle moves upwards against gravity (negative gravitaxis).\cite{Campbell2013} 
The particle orientation is described by a right-handed set of pairwise perpendicular unit vectors $\hat{u}_1$, $\hat{u}_2=\hat{u}_3\times\hat{u}_1$, and $\hat{u}_3$, where the latter is parallel to the vector pointing from the center of the platinum cap to the center of the opposing hemisphere. For an ideal particle as illustrated here, the orientation vector $\hat{u}_3$ denotes the symmetry axis of the particle and $\vec{v}_{\mathrm{A}}$ is given by $\vec{v}_{\mathrm{A}}=v_0\hat{u}_3$ with the translational self-propulsion speed $v_0=\norm{\vec{v}_{\mathrm{A}}}$.}
\end{figure}

Nano- and micron-scale self-propulsive (also called ``active'') devices\cite{Ebbens2010b,Romanczuk2012,Wang2013b,Elgeti2015,BechingerdLLRVV2016} represent a new and exciting area of research and are envisaged to be useful in a variety of applications ranging from medicine, where they could be used in targeted drug delivery and minimally invasive surgery\cite{Xi2013}, to transport of materials in microfluidic setups.\cite{Baraban2012SM,Garcia2013,Koumakis2013} 
To be able to engineer devices that have both autonomous propulsion and directional control is seen as essential to enabling many of these future applications. Whilst directional control still remains a target of research, autonomous propulsion is often achieved through the asymmetric decomposition of a dissolved fuel by an anisotropically distributed catalyst. One of the most frequently studied synthetic self-propulsive systems is fabricated by evaporating a catalyst material onto one hemisphere of a colloidal sphere to form a Janus sphere (see Fig.\ \ref{fig:ParticleSchematic}).
Typically, platinum metal is used as the catalytic material in combination with hydrogen peroxide as the dissolved fuel.\cite{Howse2007,BrownP2014} Some of us have recently shown that evaporation results in a catalytic layer that varies in thickness continuously across the surface of an insulator-sphere self-propulsive device\cite{Campbell2013} and that the dominant propulsion force\cite{tenHagenWTKBL2015,Ao2015} arises from an electrokinetic mechanism,\cite{Ebbens2014} similar as for platinum-gold nanorods.\cite{Paxton2004,Paxton2005} The asymmetrically distributed catalyst decomposes the hydrogen peroxide to water and oxygen (see Fig.\ \ref{fig:ParticleSchematic}), generating an electric field from the flow of charged intermediate species such as $\mathrm{H}^{+}$ ions across the catalyst.\cite{Ebbens2014} This results in propulsion of the Janus sphere in the direction parallel to the vector that points from the center of the metal cap to the center of the other hemisphere of the particle (see Fig.\ \ref{fig:ParticleSchematic}).\cite{Ebbens2011} 
These self-propulsive Janus colloids have been the subject of previous investigations, demonstrating promising capabilities including cargo transport directed by external magnetic fields\cite{Baraban2012}, rectifying microfluidic channels\cite{Ma2015}, or proposed methods based on chemotaxis-type effects.\cite{Ebbens2012b,Baraban2013}

Recently we extended these previous investigations, which had mainly focused on two-dimensional (2D) behavior, to consider three-dimensional (3D) trajectories of self-propulsive colloids in bulk solutions. We demonstrated that depending on the thickness of the platinum coating and on the colloid size, some trajectories displayed features similar to those observed for bottom-heavy algae cells: a gravitational alignment of the particle orientation resulting in self-propelled motion upwards towards the top of the particle's container.\cite{Campbell2013} This behavior originated because the mass of the platinum cap was biasing the orientation of the Janus sphere towards small tilt angles $\theta\in[0,\pi]$ where the heavy catalytic cap faces down (see Fig.\ \ref{fig:ParticleSchematic}). 
In this initial report of mass-anisotropy-induced gravitaxis of self-propelling devices, we only considered the case where the colloids were producing pure translational velocity. Under this simplifying assumption, the observed trajectories were explained by modeling the effect of a gravitational torque on the self-propulsion velocity vector that is otherwise randomized by Brownian rotations. However, it has been shown that self-propulsive Janus colloids often produce rotational propulsion in addition to translations. Such a rotational propulsion can be due to an accidental or engineered-in asymmetry of the platinum cap\cite{Howse2007,Archer2015} and is also observed in self-assembled agglomerates of self-propelled colloids.\cite{Ebbens2010a} The 3D trajectories of mass-anisotropic colloids with both translational and rotational self-propulsion have not yet been considered and can be expected to be complex, as three different factors can now affect device orientation: self-propulsive rotations, a gravitational torque due to an anisotropic mass distribution, and stochastic Brownian rotations.

In this work we consequently focus on experimentally and theoretically determining the 3D behavior for a range of differently sized self-propelled colloidal Janus particles for which varying amounts of gravitational alignment are expected. Particle tracking in 3D is used to measure experimental trajectories, and appropriate Langevin equations\cite{tenHagen2011JPCM,WittkowskiL2012} are derived to confirm, explain, and complement our experimental findings by theoretical considerations. 
We show that the typical trajectory of a self-propelled Janus particle is a helical path with superimposed fluctuations due to Brownian noise.\cite{Frey2005} If the mass anisotropy of the particle is sufficiently large, the helical trajectory aligns parallel or antiparallel to the direction of gravity, i.e., the particle shows gravitaxis. In case of a small mass anisotropy, however, rotational diffusion prevents gravitational alignment of the particle motion. This gravitational alignment mechanism enables the separation of self-propelled colloidal particles with respect to their mass anisotropy.
In addition, we show that angular self-propulsion has a strong influence on the shape of the trajectory. If angular self-propulsion is sufficiently strong and rotates the particle about an axis not parallel to the direction of translational self-propulsion, the trajectory has loops in quick succession. In contrast, a small angular self-propulsion leads to a more elongated or even straight trajectory.    
We demonstrate both by experiments and simulations that, as a consequence, mixtures of self-propulsive colloids with varying angular self-propulsion undergo rapid stratification. This finding allows the separation of self-propelled colloids according to their angular self-propulsion velocity.

This article is organized as follows: in Sec.\ \ref{sec:methods} we describe the experimental and theoretical methods we used to investigate the trajectories of mass-anisotropic self-propelled colloidal Janus particles. The results of our experiments and theoretical considerations are discussed in Sec.\ \ref{sec:results}. Finally, we conclude in Sec.\ \ref{sec:conclusions}.

\section{\label{sec:methods}Methods}
To study the 3D trajectories of self-propelled colloidal Janus particles, we applied both experimental and theoretical methods. 
This section addresses the setup and performance of our experiments as well as the basic equations underlying our theoretical considerations.

\subsection{\label{sec:experiments}Experimental methods}
In our experiments we prepared and tracked self-propelled Janus spheres with three different radii $R_1 = \unit[0.95]{\micro\metre}$, $R_2=\unit[1.55]{\micro\metre}$, and $R_3=\unit[2.40]{\micro\metre}$.

\subsubsection{Materials}
We purchased $\mathrm{H}_{2}\mathrm{O}_{2}$ puriss grade ($\unit[30]{\%}$ mass fraction in water) and $\unit[0.25]{mm}$ thick platinum wire ($\unit[99.99]{\%}$ purity) from Sigma Aldrich. For preparing the Janus spheres we used green fluorescent Fluoro-Max polystyrene microspheres from Thermo Scientific with radii $R_1$, 
$R_2$, and $R_3$. Furthermore, we used water from an Elga Purelab Option filtration system with electrical resistivity $\unit[0.15]{M\Omega\metre}$. 

\subsubsection{Preparation of the Janus-particle dispersion}
To prepare our catalytic Janus spheres, we followed the method described in Ref.\ [\onlinecite{Campbell2013}]. We first placed a few drops of a $\unit[0.1]{\%}$ mass fraction dispersion of green fluorescent polystyrene microspheres in ethanol on clean glass microscope slides and created a monolayer of these microspheres by spin coating. Afterwards, we evaporated a $\unit[10]{nm}$ thick layer of platinum onto the upper hemisphere of the particles. Evaporation of the platinum was done under vacuum using a Moorfield (UK) Minilab 80 electron beam evaporator. The Janus spheres were then resuspended in water by dragging the edge of a $\unit[1]{cm^2}$ square piece of moistened lens tissue repeatedly across the surface of the slide, immersing it in $\unit[1.2]{ml}$ of water in a vial, and shaking vigorously. To $\unit[1]{ml}$ of this dispersion we added $\unit[1]{ml}$ of a $\unit[30]{\%}$ mass fraction solution of $\mathrm{H}_{2}\mathrm{O}_{2}$ and we sonicated the resulting dispersion for $\unit[5]{min}$. We waited a further $\unit[25]{min}$ before adding an additional $\unit[1]{ml}$ of water to the dispersion to form a $\unit[10]{\%}$ mass fraction solution of $\mathrm{H}_{2}\mathrm{O}_{2}$. During this additional period we have observed that the platinum catalyst becomes fully activated and the particles reach their maximum self-propulsion speed. We have previously estimated the final particle volume fraction of the Janus-particle suspension to be about $\unit[0.003]{\%}$.\cite{Campbell2013} This low particle concentration has two benefits. Firstly, it avoids flows in the $\mathrm{H}_{2}\mathrm{O}_{2}$ solution that can be induced at higher particle concentrations. Secondly, the rate of catalytic decomposition of the $\mathrm{H}_{2}\mathrm{O}_{2}$ by the Janus particles is negligible over the period of the experiments: Janus particles were observed to still self-propel $\unit[24]{h}$ later with little change in speed.

\subsubsection{\label{subsubsec:Tracking}Tracking Janus-particle trajectories}
All observations of self-propelled Janus particles were made using an oxygen plasma cleaned cuvette ($\unit[40]{mm}\times\unit[10]{mm}\times\unit[0.7]{mm}$ inner dimensions, $\unit[1.25]{mm}$ wall thickness) filled with the Janus-particle dispersion. To track the self-propelled particles in three spatial dimensions, the cuvette was mounted on the stage of a Nikon Eclipse LV100 microscope, with the shortest edges of the cuvette parallel to the direction of gravity. The microscope was fitted with a Nikon $\unit[20]{x}$, $\unit[0.45]{N.A.}$ objective and operated in fluorescence mode, illuminating the Janus-particle dispersion with the blue excitation band of a Nikon B2A filter cube. Using an Andor Neo camera, we recorded a sequence of $1000$ images ($730\!\times\!730$ pixels) at a frequency of $\unit[33]{Hz}$ with the objective defocused above the green fluorescent Janus spheres. Defocussing in this way results in the particles being imaged as a bright diffraction ring, where the size of the ring is dependent on the magnification, the particle size, and the distance of the particles from the focal plane of the objective.\cite{Peterson2008} We used a set of Labview algorithms to analyze the images to extract the $(x,y)$ coordinates from the position of the ring center and the $z$ coordinate from the ring radius (see Ref.\ [\onlinecite{Campbell2013}] for details). The intensity of the fluorescence light detected by the camera falls with increasing distance of the particles from the focal plane of the objective. When the Janus particles are very far from the focal plane, the signal-to-noise ratio is too low to accurately track them.\cite{Peterson2008} As a consequence, we were limited to tracking the Janus spheres with radius $R_1$ over a maximum $z$ distance of about $\unit[100]{\micro\metre}$ and the larger particles with radius $R_3$ over $\unit[150]{\micro\metre}$. However, we found that for most of our self-propelled Janus particles this was not a significant limitation.

\subsection{\label{subsec:theory}Theoretical methods}
To study mass-anisotropic self-propelled colloidal Janus particles theoretically, we derived appropriate Langevin equations using the general framework described in Ref.\ [\onlinecite{WittkowskiL2012}].
This framework has already successfully been applied to calculate the $6\!\times\!6$-dimensional diffusion tensor of anisometric colloidal particles from their orientation-resolved trajectories\cite{KraftWtHEPL2013} and to predict the trajectories of an anisometric but mass-isotropic self-propelled colloidal particle in two spatial dimensions in the absence\cite{KuemmeltHWBEVLB2013,KuemmeltHWTBEVLB2014} and in the presence\cite{tenHagenKWTLB2014} of gravity. Here, we apply this framework to bottom-heavy (i.e., mass-anisotropic) self-propelled colloidal particles under gravity. In order to derive corresponding Langevin equations, we assumed a spherical particle shape and included a torque that tends to orient the particles upwards.

We describe the translation and rotation of such a particle by the time-dependent position $\vec{r}(t)$ of its center and the time-dependent orthonormal orientational unit vectors $\hat{u}_{1}(t)$, $\hat{u}_{2}(t)$, and $\hat{u}_{3}(t)=\hat{u}_{1}(t)\!\times\!\hat{u}_{2}(t)$, respectively. The orientational unit vectors rotate with the particle and are chosen in such a way that $\hat{u}_{3}(t)$ is always parallel to the vector that points from the center of the metal-coated hemisphere to the center of the opposing hemisphere of the particle. For an ideal Janus particle, which is rotationally symmetric about $\hat{u}_{3}(t)$, this unit vector thus denotes the direction of the translational self-propulsion at time $t$. 
A real Janus particle, however, is not ideal and small imperfections in the particle shape and catalytic layer cause a deviation of the direction of translational self-propulsion from the particle orientation vector $\hat{u}_{3}(t)$. For such a particle, the translational self-propulsion velocity can be written as $\vec{v}_{\mathrm{A}}=v_{\mathrm{A},1}\hat{u}_{1}+v_{\mathrm{A},2}\hat{u}_{2}+v_{\mathrm{A},3}\hat{u}_{3}$ with constant coefficients $v_{\mathrm{A},1}$, $v_{\mathrm{A},2}$, and $v_{\mathrm{A},3}$. In most situations one has $v_{\mathrm{A},3}\gg v_{\mathrm{A},1}$ and $v_{\mathrm{A},3}\gg v_{\mathrm{A},2}$. Analogously, the angular self-propulsion velocity can be written as $\vec{\omega}_{\mathrm{A}}=\omega_{\mathrm{A},1}\hat{u}_{1}+\omega_{\mathrm{A},2}\hat{u}_{2}+\omega_{\mathrm{A},3}\hat{u}_{3}$ with constant coefficients $\omega_{\mathrm{A},1}$, $\omega_{\mathrm{A},2}$, and $\omega_{\mathrm{A},3}$.    
Besides the self-propulsion of the particle, we take into account the effect of gravity on the particle motion. This is achieved by means of a sedimentation velocity $-v_{\mathrm{s}}\hat{e}_{z}$\cite{Tailleur2009,Palacci2010,Wolff2013,Ginot2015,Xiao2015,Chen2016} with sedimentation speed $v_{\mathrm{s}}$ and the unit vector parallel to the $z$ axis $\hat{e}_{z}=(0,0,1)^\mathrm{T}$ as well as through an orientation-dependent angular velocity $\omega_{\mathrm{M}}\vec{a}$ that tends to rotate the particle upwards so that $\hat{u}_{3}$ becomes identical to $\hat{e}_{z}$. This angular velocity with maximum value $\omega_{\mathrm{M}}\geqslant 0$ and with the vector $\vec{a}=(u_{3,y},-u_{3,x},0)^{\mathrm{T}}$, which depends on the orientation vector $\hat{u}_{3}=(u_{3,x},u_{3,y},u_{3,z})^{\mathrm{T}}$, corresponds to the aligning torque that results from the displacement of the center of mass of the Janus particle towards the relatively heavy metal cap (see Fig.\ \ref{fig:ParticleSchematic}). 

Appropriate Langevin equations for describing the motion of such a particle are given by
{\allowdisplaybreaks\begin{gather}%
\dot{\vec{r}} = v_{\mathrm{A},1}\hat{u}_{1} + v_{\mathrm{A},2}\hat{u}_{2} + v_{\mathrm{A},3}\hat{u}_{3} - v_{\mathrm{s}}\hat{e}_{z} + \sqrt{2D_{\mathrm{T}}}\vec{\xi}_{\mathrm{T}} \,,\label{eq:LGa}\\
\vec{\omega} = \omega_{\mathrm{A},1}\hat{u}_{1} + \omega_{\mathrm{A},2}\hat{u}_{2} + \omega_{\mathrm{A},3}\hat{u}_{3} + \omega_{\mathrm{M}}\vec{a}
+\sqrt{2D_{\mathrm{R}}}\vec{\xi}_{\mathrm{R}} \,,\label{eq:LGb}\\
\dot{\hat{u}}_{i} = \vec{\omega}\!\times\!\hat{u}_{i} \quad\text{for}\quad i\in\{1,2,3\}
\label{eq:LGc}%
\end{gather}}%
with the translational particle velocity $\dot{\vec{r}}(t)=\dif\vec{r}(t)/\dif t$, the angular particle velocity $\vec{\omega}(t)$, and the translational and rotational diffusion coefficients $D_{\mathrm{T}}=k_{\mathrm{B}}T/(6\pi\eta R)$ and $D_{\mathrm{R}}=3D_{\mathrm{T}}/(4R^2)$, respectively, where $k_{\mathrm{B}}$ is the Boltzmann constant, $T$ is the absolute temperature, $\eta$ is the dynamic viscosity of the $\mathrm{H}_{2}\mathrm{O}_{2}$ solution surrounding the Janus particles, and $R$ is the particle radius. 
$\vec{\xi}_{\mathrm{T}}(t)$ and $\vec{\xi}_{\mathrm{R}}(t)$ model the Brownian noise that affects the particle motion. Their individual components are statistically independent, zero-mean, unit-variance Gaussian white noises.

In order to simulate trajectories of mass-anisotropic self-propelled Janus particles, we solved the Langevin equations \eqref{eq:LGa}-\eqref{eq:LGc} numerically using the Euler-Maruyama method\cite{KloedenP2006}. For these numerical calculations, we chose parameters that correspond to our experiments. 
This means that we considered the three particle radii $R_1 = \unit[0.95]{\micro\metre}$, $R_2=\unit[1.55]{\micro\metre}$, and $R_3=\unit[2.40]{\micro\metre}$, the temperature $T=\unit[293]{\kelvin}$, and  the viscosity $\eta=1.02\cdot\unit[10^{-3}]{\pascal\,\second}$. 
Furthermore, depending on the particle radius $R\in\{R_1,R_2,R_3\}$ we chose the translational self-propulsion speed $v_0=\norm{\vec{v}_{\mathrm{A}}}=\sqrt{v^{2}_{\mathrm{A},1}+v^{2}_{\mathrm{A},2}+v^{2}_{\mathrm{A},3}}$, the sedimentation speed $v_{\mathrm{s}}$, the angular self-propulsion speed $\omega_0=\norm{\vec{\omega}_{\mathrm{A}}}=\sqrt{\omega^{2}_{\mathrm{A},1}+\omega^{2}_{\mathrm{A},2}+\omega^{2}_{\mathrm{A},3}}$, and the maximal aligning angular velocity $\omega_{\mathrm{M}}$ as shown in Table \ref{tab:Parameter}.
\begin{table}[htb]
\centering
\begin{ruledtabular}
\begin{tabular}{cccccc}%
$R/\unit{\micro\metre}$ & $v_0/(\unit{\micro\metre\,\second^{-1}})$ & $\mathrm{Pe}$ & $v_{\mathrm{s}}/(\unit{\micro\metre\,\second^{-1}})$ & $\omega_0/\unit{\second^{-1}}$ & $\omega_{\mathrm{M}}/\unit{\second^{-1}}$ \\
\hline
$0.95$ & $5.33$ & $45.7$ & $0.11$ & $[0, 1.72]$ & $0.053$ \\
$1.55$ & $4.35$ & $99.3$ & $0.32$ & $[0, 1.87]$ & $0.093$ \\
$2.40$ & $4.18$ & $229$ & $0.50$ & $[0, 0.79]$ & $0.082$ \\
\end{tabular}
\end{ruledtabular}
\caption{\label{tab:Parameter}Parameters $v_0$, $\mathrm{Pe}$, $v_{\mathrm{s}}$, $\omega_0$, and $\omega_{\mathrm{M}}$ for our Janus particles with different radii $R$ and a $\unit[10]{nm}$ thick platinum cap.}%
\end{table}
The values for $v_0$ were directly measured in our experiments. As can be seen from the corresponding values of the P\'eclet number $\mathrm{Pe}=2R v_{0}/D_{\mathrm{T}}$ (see Table \ref{tab:Parameter}), these translational self-propulsion speeds are quite large so that for all considered particles self-propulsion clearly dominates passive diffusion. We measured the values for $v_{\mathrm{s}}$ in water, where the particles are not self-propelled, and deduced from these measurements the values of $v_{\mathrm{s}}$ in the $\mathrm{H}_{2}\mathrm{O}_{2}$ solution. The values for $\omega_{\mathrm{M}}$ were calculated from the mass distribution of the Janus particles with a $\unit[10]{nm}$ thick platinum cap, using the relation $\vec{\omega}=(D_{\mathrm{R}}/(k_{\mathrm{B}}T))\vec{M}$ between a torque $\vec{M}$ acting on a spherical colloidal particle and its resulting angular velocity $\vec{\omega}$. (See the Appendix for details on the calculation of $v_{\mathrm{s}}$ and $\omega_{\mathrm{M}}$.) 
However, it was neither possible to measure $\omega_0$ directly, which would require to track the particles with respect to all three orientational degrees of freedom, nor to calculate it from other known quantities. 

Therefore, we estimated $\omega_0$ from the translational motion of the particles tracked in our experiments. For this purpose, we neglected the $z$ coordinates of the particle positions, which are biased by gravitaxis, and calculated the angles $\Delta\phi_i\in(-\pi,\pi)$ by which the direction of translational motion of a particle changed when proceeding from the $i$th to the $(i+1)$th time step.\footnote{The values $-\pi$ and $\pi$ are excluded here from the domain of the angles $\Delta\phi_i$, since they cannot be distinguished. However, in the experiments we typically observed angles $|\Delta\phi_i|\ll 1$ anyway.} 
>From these angles we in turn determined the angular speed $\omega_{\mathrm{xy}}=(N-1)^{-1}\abs{\sum^{N-1}_{i=1}\Delta\phi_i/\Delta t}$, where $N$ is the number of time steps and $\Delta t$ is the time step size of the considered trajectory. The resulting mean values $\langle\omega_{\mathrm{xy}}\rangle$ are $\langle\omega_{\mathrm{xy}}\rangle_{R_1}=\unit[0.860]{\second^{-1}}$, $\langle\omega_{\mathrm{xy}}\rangle_{R_2}=\unit[0.933]{\second^{-1}}$, and $\langle\omega_{\mathrm{xy}}\rangle_{R_3}=\unit[0.396]{\second^{-1}}$ for particles of radii $R_1$, $R_2$, and $R_3$, respectively. We therefore used the intervals $[0,2\langle\omega_{\mathrm{xy}}\rangle_{R_i}]$ to approximate the values of $\omega_0$ for particles of radius $R_{i}$ with $i\in\{1,2,3\}$ (see Table \ref{tab:Parameter}). 
When simulating a trajectory for a Janus particle using the Langevin equations \eqref{eq:LGa}-\eqref{eq:LGc}, we selected a value for $\omega_0$ randomly and with uniform probability from the corresponding interval in Table \ref{tab:Parameter}. 

After determining values for $v_0$ and $\omega_0$, we used them to choose reasonable values for the parameters $v_{\mathrm{A},i}$ and $\omega_{\mathrm{A},i}$ with $i\in\{1,2,3\}$.
Since for the Janus particles studied in this article the translational self-propulsion acts mainly in the direction of $\hat{u}_{3}$ and the angular self-propulsion velocity is assumed to have a component with a small amount parallel to $\hat{u}_{3}$, we chose values for $v_{\mathrm{A},1}$ and $v_{\mathrm{A},2}$ randomly and with uniform probability from the interval $[-0.1v_0,0.1v_0]$ and for $\omega_{\mathrm{A},1}$ and $\omega_{\mathrm{A},2}$ from the set $[-\sqrt{0.5}\omega_0,-\sqrt{0.45}\omega_0]\cup[\sqrt{0.45}\omega_0,\sqrt{0.5}\omega_0]$.
The values for $v_{\mathrm{A},3}$ and $\omega_{\mathrm{A},3}$ then followed from the conditions $v_0=\norm{\vec{v}_{\mathrm{A}}}$ and $\omega_0=\norm{\vec{\omega}_{\mathrm{A}}}$, where a random sign had to be chosen for $\omega_{\mathrm{A},3}$. 

In our simulations of trajectories of self-propelled Janus particles we used the origin of coordinates $\vec{r}(0)=(0,0,0)^{\mathrm{T}}$ and a random orientation described by randomly chosen and pairwise perpendicular unit vectors $\hat{u}_{1}(0)$, $\hat{u}_{2}(0)$, and $\hat{u}_{3}(0)=\hat{u}_{1}(0)\times\hat{u}_{2}(0)$ as initial conditions. We simulated the trajectories for a period of length $\unit[999/33]{\second}\approx\unit[30]{\second}$, which corresponds to recording $1000$ images at frame rate $\unit[33]{Hz}$ in the experiments. For better numerical accuracy, we solved the Langevin equations \eqref{eq:LGa}-\eqref{eq:LGc} with a time step size $\Delta t/4$, which is smaller by a factor of $4$ than the time steps in our experiments, but for analogy with the experimental data we downsampled the resulting trajectories to the time step size $\Delta t$.

\section{\label{sec:results}Results and discussion}
\begin{figure*}[htb]
\centering
\includegraphics[width=\linewidth]{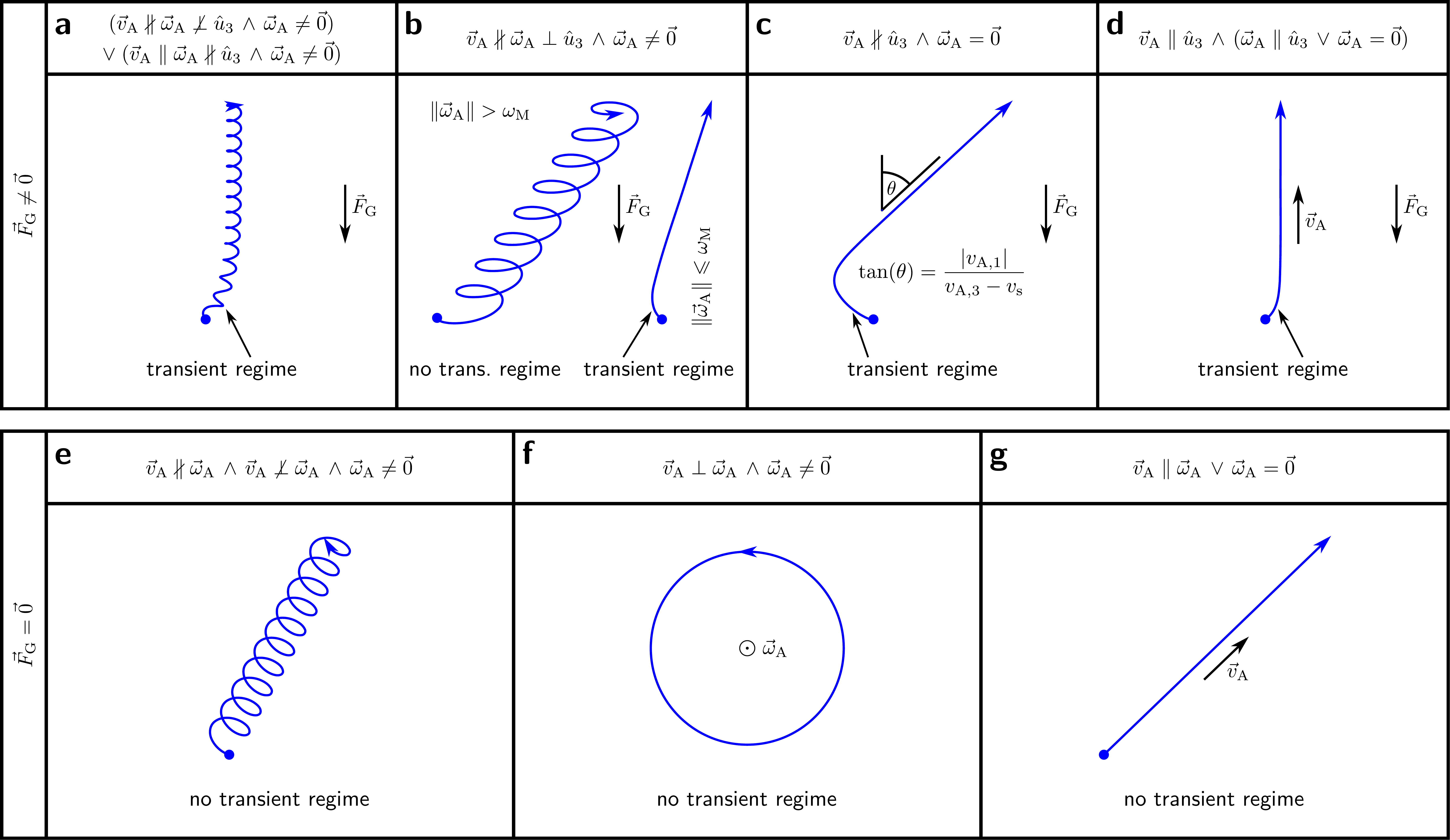}%
\caption{\label{fig:TrajectoriesTh}Classification of the noise-free solutions of the Langevin equations \eqref{eq:LGa}-\eqref{eq:LGc}. Depending on the vectors $\vec{v}_{\mathrm{A}}=v_{\mathrm{A},1}\hat{u}_{1}+v_{\mathrm{A},2}\hat{u}_{2}+v_{\mathrm{A},3}\hat{u}_{3}$ and $\vec{\omega}_{\mathrm{A}}=\omega_{\mathrm{A},1}\hat{u}_{1}+\omega_{\mathrm{A},2}\hat{u}_{2}+\omega_{\mathrm{A},3}\hat{u}_{3}$, which model the self-propulsion of the particles, and the gravitational force $\vec{F}_{\mathrm{G}}$, qualitatively different trajectories can be observed. Note that the trajectories shown in (a)-(d) point downwards if the translational self-propulsion is weak.}%
\end{figure*}

First, we study the qualitative shapes of the trajectories of the mass-anisotropic self-propelled colloidal Janus particles, which depend on the values of the parameters $v_{\mathrm{A},i}$ and $\omega_{\mathrm{A},i}$ with $i\in\{1,2,3\}$ and thus on the particular shapes and sizes of these non-ideal particles and their catalytic layers. 
For this purpose we analyzed the Langevin equations \eqref{eq:LGa}-\eqref{eq:LGc} and numerically solved them for various different parameter choices and random initial particle orientations.
Regarding the parameters $v_{\mathrm{A},i}$ and $\omega_{\mathrm{A},i}$, we only stipulated $v_{\mathrm{A},3}>0$ to focus on forwards moving particles, which is in accordance with our experiments. Apart from that, we allowed for all real values of these parameters. 
Since we are interested only in the qualitative shapes of the trajectories, we neglected the noise terms in Eqs.\ \eqref{eq:LGa}-\eqref{eq:LGc}. Furthermore, we chose $v_{\mathrm{s}}/\omega_{\mathrm{M}}$ and $1/\omega_{\mathrm{M}}$ as units for distance and time, respectively, so that the equations depend only on the dimensionless ratios $v_{\mathrm{A},i}/v_{\mathrm{s}}$ and $\omega_{\mathrm{A},i}/\omega_{\mathrm{M}}$, but no longer explicitly on $v_{\mathrm{s}}$ and $\omega_{\mathrm{M}}$.  
The qualitative shapes of the trajectories can then be classified with respect to the relative orientations of the translational self-propulsion velocity $\vec{v}_{\mathrm{A}}$, angular self-propulsion velocity $\vec{\omega}_{\mathrm{A}}$, and orientation of the particle axis $\hat{u}_{3}$\footnote{Note that for constant parameters $v_{\mathrm{A},i}$ and $\omega_{\mathrm{A},i}$ with $i\in\{1,2,3\}$ the relative orientations of the vectors $\vec{v}_{\mathrm{A}}$, $\vec{\omega}_{\mathrm{A}}$, and $\hat{u}_{3}$ are time-independent.} and by distinguishing the cases $\vec{\omega}_{\mathrm{A}}\neq\vec{0}$ and $\vec{\omega}_{\mathrm{A}}=\vec{0}$.
Figure \ref{fig:TrajectoriesTh} shows such a classification of the possible trajectory shapes of mass-anisotropic self-propelled colloidal Janus particles. 
Remarkably, the typical trajectory of a self-propelled Janus particle is a helical path. This is, of course, superimposed by fluctuations when Brownian noise is taken into account.

If gravity is present ($\vec{F}_{\mathrm{G}}\neq\vec{0}$), the trajectories typically start with an irregular transient regime, which depends on the initial particle orientation, and then converge into a helical curve pointing in vertical direction so that the axis of the helix is parallel or antiparallel to the buoyancy-corrected gravitational force $\vec{F}_{\mathrm{G}}$ (see Fig.\ \ref{fig:TrajectoriesTh}a). This implies that the third component of the angular velocity $\vec{\omega}$ is constant. While the $z$ coordinate of the particle's center-of-mass position increases linearly as soon as the particle enters the periodic regime, the projection onto the $x$-$y$ plane is a circle. Moreover, the tilting angle of the particle with respect to the direction of gravity is constant. This means that only the first two components of the particle's orientation vector $\hat{u}_{3}$ change in time. Such trajectories are similar to those of self-propelled colloidal particles under gravity that have a homogeneous mass distribution but an anisotropic shape \cite{WittkowskiL2012}.
There are only three exceptions from the typical helical trajectories. These special cases can occur if $\vec{\omega}_{\mathrm{A}}$ is parallel or perpendicular to $\hat{u}_{3}$ or if $\vec{\omega}_{\mathrm{A}}$ vanishes. 

The first exception is realized if $\vec{\omega}_{\mathrm{A}}$ is perpendicular to $\hat{u}_{3}$, not parallel to $\vec{v}_{\mathrm{A}}$, and not vanishing.
In this special case there is a direct competition between the angular self-propulsion velocity $\vec{\omega}_{\mathrm{A}}$ and the aligning torque due to gravity so that depending on the modulus of $\vec{\omega}_{\mathrm{A}}$ two qualitatively different trajectories can be observed (see Fig.\ \ref{fig:TrajectoriesTh}b). 
As long as the maximal aligning torque occurring for $\hat{u}_{3}\perp\vec{F}_{\mathrm{G}}$ is larger than the torque that corresponds to the angular self-propulsion velocity $\vec{\omega}_{\mathrm{A}}$, a particle starts with an initial transient regime and then proceeds with a straight tilted trajectory. Since the torques acting on the particle and the corresponding angular velocities differ only by a factor $D_{\mathrm{R}}/(k_{\mathrm{B}}T)$, the condition for this trajectory can be expressed by $\omega_0\leqslant\omega_{\mathrm{M}}$. 
On the other hand, if $\omega_0$ exceeds $\omega_{\mathrm{M}}$, the particle moves on a circling path without any transient regime. This path is periodic but not a helix, with an elliptical instead of a circular cross section, and its orientation depends on the initial orientation of the particle. Such a transition from straight to circling periodic trajectories, occurring if the angular self-propulsion exceeds a critical value, is similar to what has been observed for asymmetric self-propelled colloidal particles in two spatial dimensions with a homogeneous mass distribution under gravity \cite{tenHagenKWTLB2014}. 
The second exception is given if $\vec{\omega}_{\mathrm{A}}$ vanishes and $\vec{v}_{\mathrm{A}}$ is not parallel to $\hat{u}_{3}$. Then the trajectory is again a tilted straight line after an initial transient regime (see Fig.\ \ref{fig:TrajectoriesTh}c) with the tilting angle $\theta$ between the straight part of the trajectory and $\vec{F}_{\mathrm{G}}$ now given by $\tan(\theta)=\abs{v_{\mathrm{A},1}}/(v_{\mathrm{A},3}-v_{\mathrm{s}})$. 
Finally, the third exception requires that $\vec{\omega}_{\mathrm{A}}$ is either vanishing or parallel to $\hat{u}_{3}$ and that $\vec{v}_{\mathrm{A}}$ is parallel to $\hat{u}_{3}$.
In this special situation, the particle starts with an initial transient regime and converges into a straight line that is parallel to $\vec{v}_{\mathrm{A}}$ and parallel or antiparallel to $\vec{F}_{\mathrm{G}}$ (see Fig.\ \ref{fig:TrajectoriesTh}d).
The reason for this vertical alignment of the trajectory is that the particle is now rotationally symmetric about $\hat{u}_{3}$ and that gravity can rotate the particle so that its heavy platinum cap is pointing downwards and $\hat{u}_{3}$ becomes antiparallel to $\vec{F}_{\mathrm{G}}$.

This means that for a sufficiently strong translational self-propulsion that dominates gravity and thus avoids sedimentation, the particles move upwards under gravity, i.e., they show (negative) gravitaxis. As long as gravity is present, this is a general result and true even in all three special cases mentioned above (see Figs.\ \ref{fig:TrajectoriesTh}a-d). We will address this effect in more detail further below.   

If, in contrast, there is no gravity ($\vec{F}_{\mathrm{G}}=\vec{0}$), the trajectories of self-propelled Janus particles are much simpler, since their mass distribution becomes irrelevant. Here the general trajectory shape is a helix\cite{WittkowskiL2012,Sevilla2016} which has no transient regime and an orientation that depends on the initial particle orientation (see Fig.\ \ref{fig:TrajectoriesTh}e). Only two limiting special cases can be distinguished from this general case. The first one occurs if the angular self-propulsion velocity $\vec{\omega}_{\mathrm{A}}$ of the particle is non-vanishing and perpendicular to the translational self-propulsion velocity $\vec{v}_{\mathrm{A}}$. 
In this case, the particle can only move in a plane perpendicular to $\vec{\omega}_{\mathrm{A}}$ and the helix reduces to a circle (helix with zero pitch length) whose position and orientation depend on the initial position and orientation of the particle (see Fig.\ \ref{fig:TrajectoriesTh}f).
On the other hand, for a vanishing $\vec{\omega}_{\mathrm{A}}$ or if $\vec{\omega}_{\mathrm{A}}$ and $\vec{v}_{\mathrm{A}}$ are parallel, the particle does not rotate or the rotation has no effect on its translational motion, respectively. In this special case, the particle can only move parallel to $\vec{v}_{\mathrm{A}}$ and the trajectory is infinitely stretched to a straight line (helix with infinite pitch length) whose orientation again depends on the initial particle orientation (see Fig.\ \ref{fig:TrajectoriesTh}g).   

\begin{figure*}[htb]
\centering
\includegraphics[width=\linewidth]{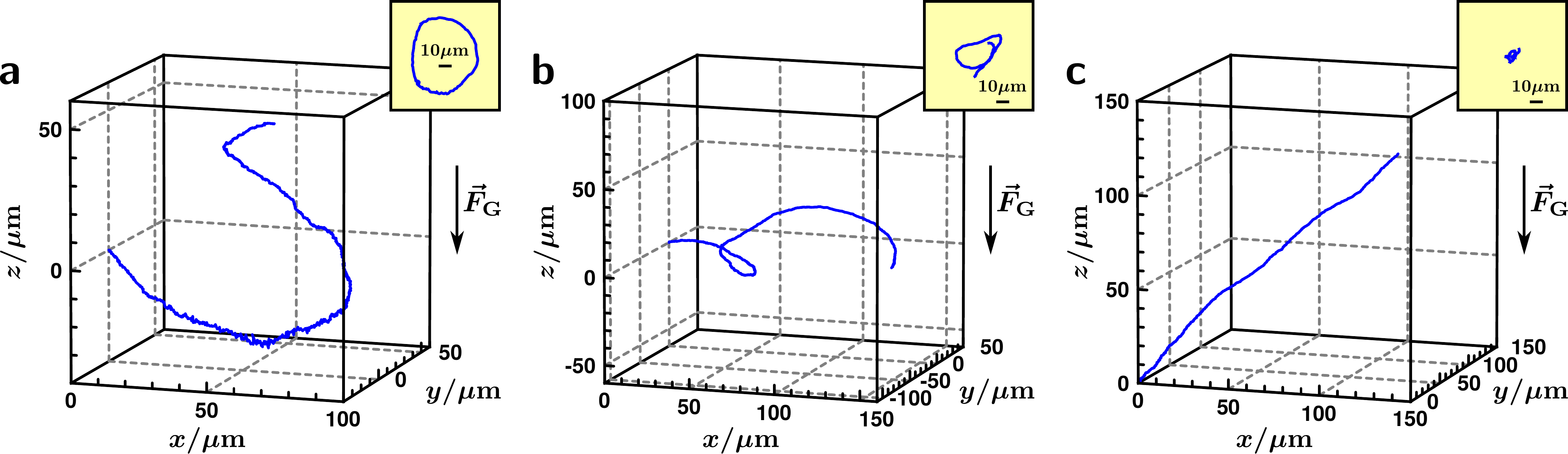}%
\caption{\label{fig:TrajectoriesExp}Different types of trajectories observed in the experiments: (a) a turn of a helix, (b) about one turn of a helix-like curve, and (c) a straight line. The insets show corresponding orthographic projections parallel to the curves' axes.}%
\end{figure*}

In our experiments, of course, not all parameter values were realized so that it was not possible to observe all the trajectory types distinguished in Fig.\ \ref{fig:TrajectoriesTh}.
The dominant trajectory types which we observed in the experiments are shown in Fig.\ \ref{fig:TrajectoriesExp}. 
They are a helical curve (see Fig.\ \ref{fig:TrajectoriesExp}a), a helix-like curve with an elliptical cross section (see Fig.\ \ref{fig:TrajectoriesExp}b), and a tilted straight line (see Fig.\ \ref{fig:TrajectoriesExp}c). These trajectories correspond to the general helical trajectories illustrated in Fig.\ \ref{fig:TrajectoriesTh}a, the non-helical periodic trajectories represented in Fig.\ \ref{fig:TrajectoriesTh}b, and the tilted straight trajectories of Figs.\ \ref{fig:TrajectoriesTh}b or \ref{fig:TrajectoriesTh}c, respectively.
A trajectory with a straight vertical upwards motion corresponding to Fig.\ \ref{fig:TrajectoriesTh}d was not observed in our experiments. In view of the fact that this case requires very special parameter combinations, it is not surprising that this situation is rare in experiments. 
Note that in the case of periodic trajectories as shown in Figs.\ \ref{fig:TrajectoriesExp}a and \ref{fig:TrajectoriesExp}b, the reasons for observing only individual loops are the limited tracking time of $\approx\unit[30]{\second}$ in the experiments and the limited $z$ distance over which particles could be tracked (see Sec.\ \ref{subsubsec:Tracking}). 

\begin{figure*}[htb]
\centering
\includegraphics[width=\linewidth]{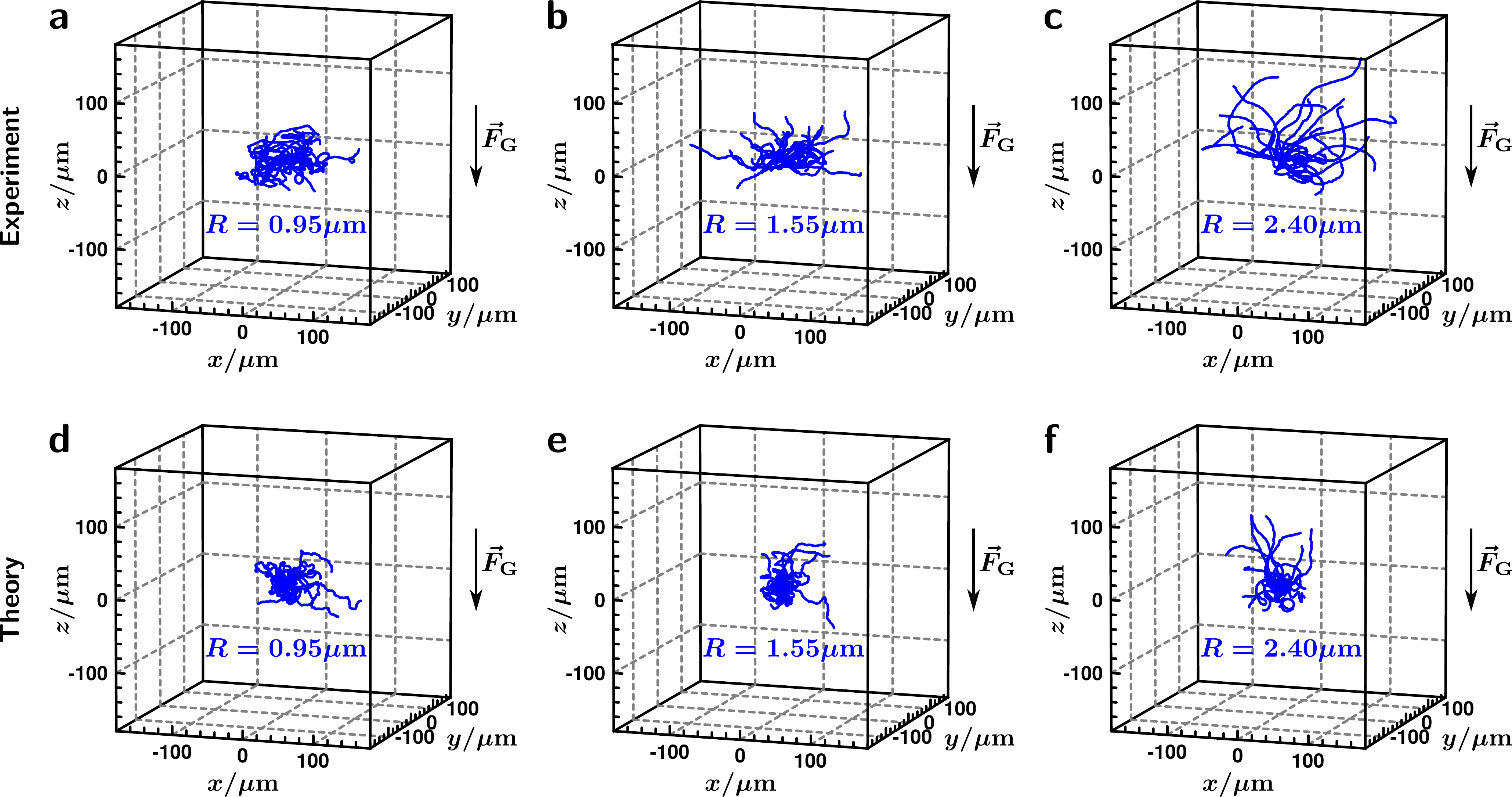}%
\caption{\label{fig:Gravitaxis}Trajectories of various Janus spheres with different radii $R$ (a)-(c) observed in our experiments and (d)-(f) simulated using Eqs.\ \eqref{eq:LGa}-\eqref{eq:LGc}. With increasing radius the particle motion is more and more biased upwards (negative gravitaxis) and the trajectories become increasingly elongated.}%
\end{figure*}

Next, we study the full trajectories of self-propelled Janus particles including Brownian noise. In Fig.\ \ref{fig:Gravitaxis} a group of trajectories starting in the origin of ordinates and corresponding to different particles is shown for each particle radius $R\in\{R_1,R_2,R_3\}$. 
The experimental results (see Figs.\ \ref{fig:Gravitaxis}a-c), which include the trajectories from Fig.\ \ref{fig:TrajectoriesExp}, show a strong dependence on the particle radius. While for the smallest particles of radius $R_1 = \unit[0.95]{\micro\metre}$ rotational diffusion is large and the trajectories appear as a cloud that is centered around the origin of ordinates, for the larger particles with radii $R_2=\unit[1.55]{\micro\metre}$ and $R_3=\unit[2.40]{\micro\metre}$ the trajectories are increasingly biased upwards, i.e., opposite to the gravitational force $\vec{F}_{\mathrm{G}}$ (negative gravitaxis).
This originates from the mass anisotropy of the particles, which is due to their platinum cap and leads to a preferred upwards orientation. The aligning effect increases with the particle radius and leads to gravitaxis when it dominates rotational diffusion. 
Furthermore, while the trajectories of the smallest particles are balled up, the trajectories of the larger particles are more and more elongated. This is in accordance with the fact that the P\'eclet number $\mathrm{Pe}=2R v_{0}/D_{\mathrm{T}}=3v_{0}/(2R D_{\mathrm{R}})$ grows with the particle radius (see Table \ref{tab:Parameter}). However, it is an effect not only of rotational diffusion, which decreases for a growing particle radius, but also of the increasing aligning torque due to the mass anisotropy of the particles.
For comparison, we simulated the same number of corresponding trajectories by numerically solving the Langevin equations \eqref{eq:LGa}-\eqref{eq:LGc} (see Sec.\ \ref{subsec:theory} for the chosen parameters). Our simulation results (see Figs.\ \ref{fig:Gravitaxis}d-f) are in good agreement with the experimental results and show the same features. 
Note that some of the experimental trajectories are longer than the corresponding simulated trajectories, since -- for each particle size -- we simulated all trajectories with the same translational self-propulsion speed $v_0$ given in Table \ref{tab:Parameter}, whereas in the experiments some particles reached larger speeds. The facts that the motion of the Janus particles becomes gravitactic if their mass anisotropy is sufficiently large and that the elongation of the trajectories increases with the particle radius are potentially useful features, since they allow to separate self-propelled colloidal Janus particles with respect to their mass anisotropy and size.
 
\begin{figure}[htb]
\centering
\includegraphics[width=\linewidth]{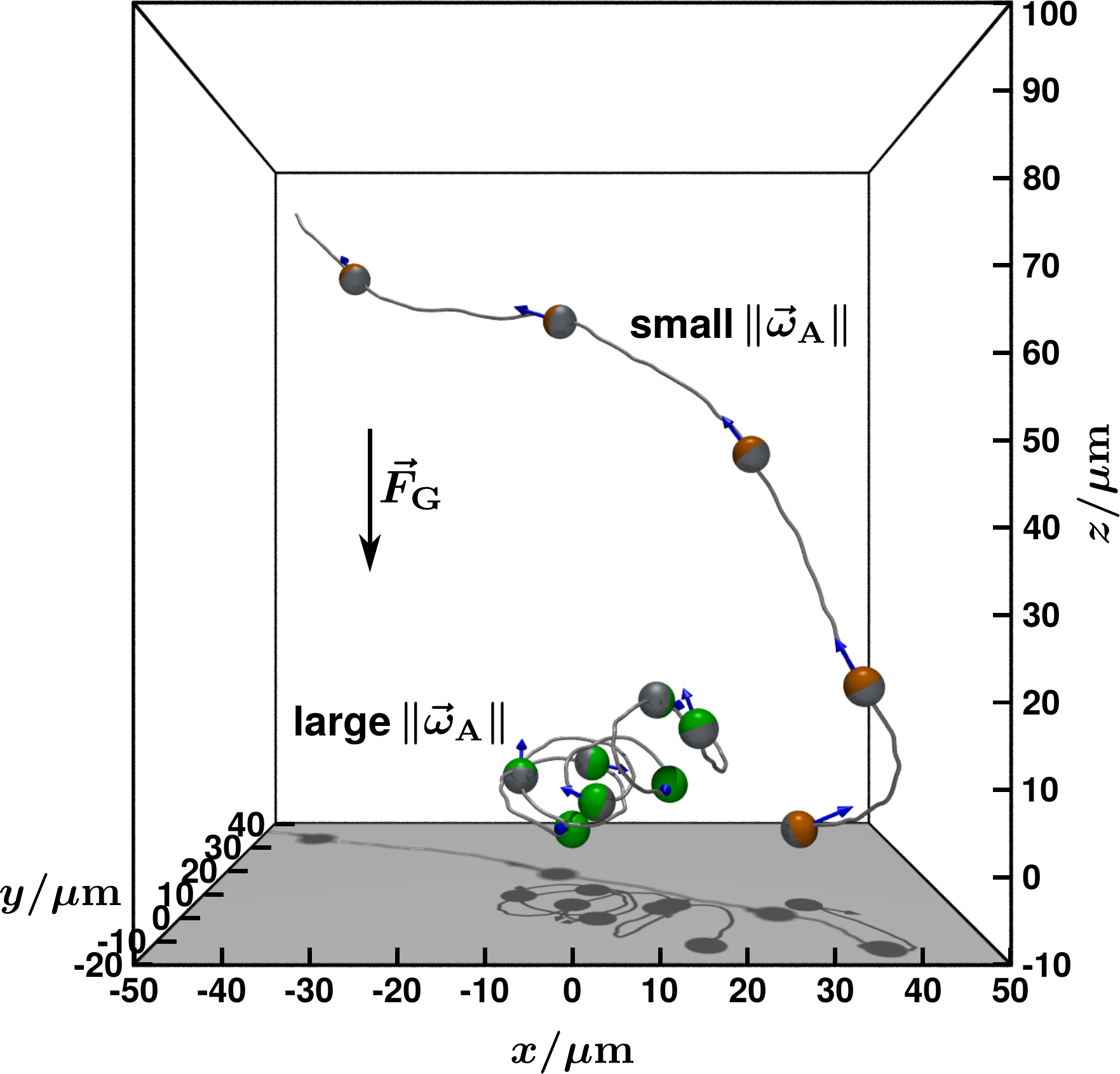}%
\caption{\label{fig:TwoTrajectories}Experimental trajectories of Janus spheres with the same radius $R_3=\unit[2.40]{\micro\metre}$ and translational self-propulsion speed $v_0$ but different angular self-propulsion speeds $\norm{\vec{\omega}_{\mathrm{A}}}$. The particle with larger $\norm{\vec{\omega}_{\mathrm{A}}}$ has a balled-up trajectory with many loops, whereas the particle with smaller $\norm{\vec{\omega}_{\mathrm{A}}}$ has a more elongated trajectory that runs upwards due to negative gravitaxis. Particle positions are shown at $\unit[5]{\second}$ intervals.}%
\end{figure}

When comparing trajectories of equally-sized self-propelled Janus particles, it is observed that even for particles with the same radius $R$ and translational self-propulsion speed $v_0$ the trajectories can have qualitatively different shapes.
Figure \ref{fig:TwoTrajectories} shows two such trajectories for particles of radius $R_3=\unit[2.40]{\micro\metre}$. 
Obviously, the trajectory of one particle has many loops and is balled up, whereas the trajectory of the other particle is much more elongated and shows less than one full loop. 
Within our modeling, this behavior of the particles can be attributed to different values of the angular self-propulsion speed $\omega_0=\norm{\vec{\omega}_{\mathrm{A}}}$, which has a strong influence on the shape of the trajectory. 
When $\omega_0$ is sufficiently large and $\vec{\omega}_{\mathrm{A}}$ rotates the particle about an axis not parallel to $\vec{v}_{\mathrm{A}}$, the trajectory has small loops in quick succession. In contrast, a small $\omega_0$ leads to a more elongated or even straight trajectory.
Differing values of $\omega_0$ in this work are thought to result from small imperfections in the shape of the particle's catalytic hemisphere, which introduce asymmetries leading to rotational propulsion.  
In a recent article, we have demonstrated that Janus spheres with large values of $\omega_0$ can be intentionally produced with a high degree of precision.\cite{Archer2015}

The strong dependence of the trajectory shape on the angular self-propulsion velocity is another interesting feature of the particles that enables them to be sorted. 
In a mixture of self-propelled colloidal particles differing only by their degree of rotational propulsion, the particles with slow rotation and elongated trajectories will move fast away from their initial position, while particles with larger angular self-propulsion and balled-up trajectories remain closer to their initial position and thus separate from the other particles.
When gravitaxis is relevant, this spatial separation effect is particularly strong in the vertical direction. As an example, the difference in climb rate of the particles in Fig.\ \ref{fig:TwoTrajectories} would result in a vertical separation distance of about $\unit[1]{mm}$ in $\unit[5]{min}$.

\begin{figure}[tb]
\centering
\includegraphics[width=\linewidth]{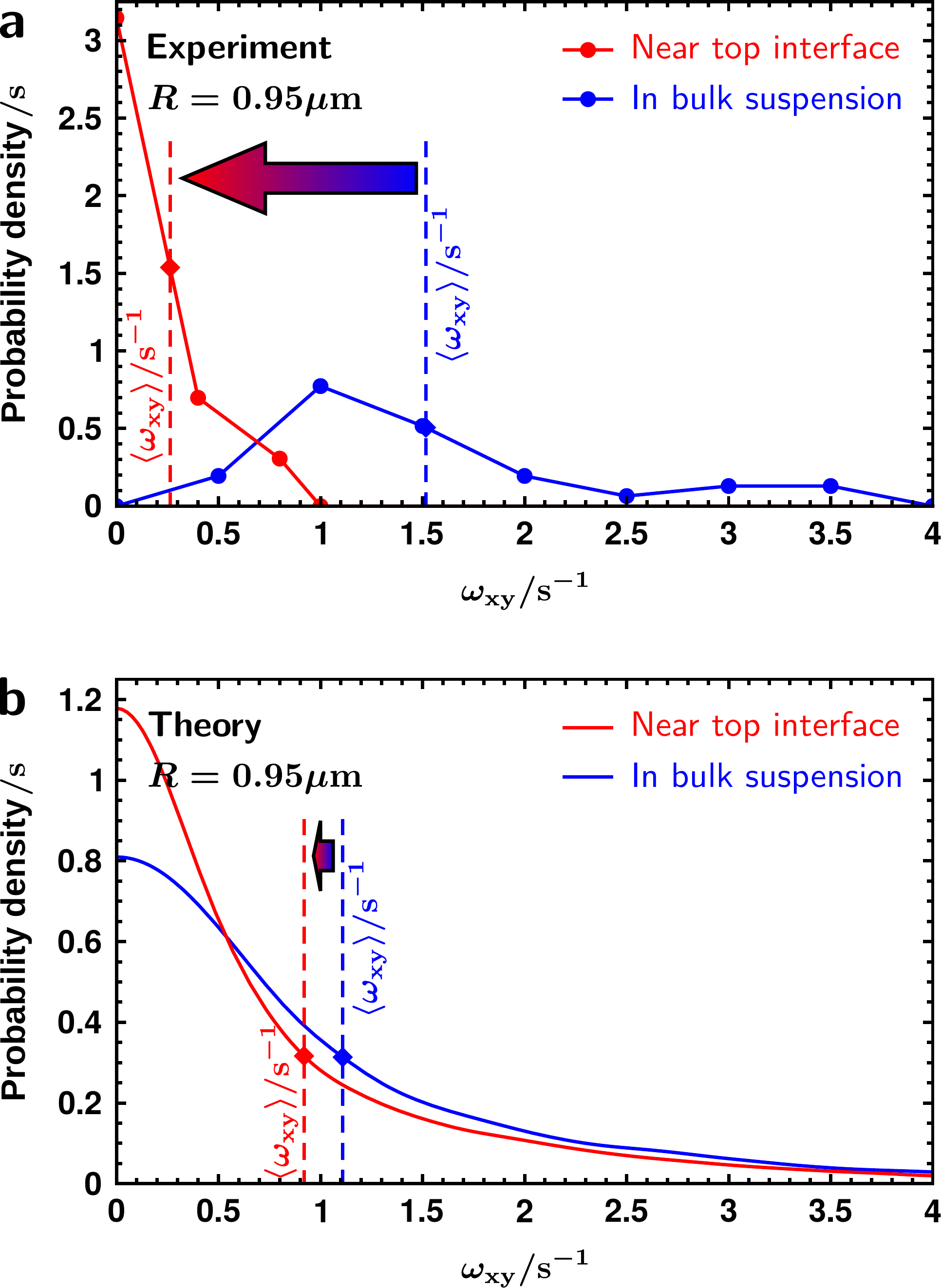}
\caption{\label{fig:Separation}(a) Experimental and (b) theoretical results for the probability distributions of the two-dimensional angular speed $\omega_{\mathrm{xy}}$ in the center (i.e., in the bulk) and near the upper interface of a suspension of self-propelled colloidal particles with radius $R_1 = \unit[0.95]{\micro\metre}$ and a $\unit[20]{nm}$ thick platinum coating in a cuvette with inner height $\unit[700]{\micro\metre}$. A shift of the expected values $\langle\omega_{\mathrm{xy}}\rangle$ of the probability distributions near the top interface to smaller values than in the bulk suspension is clearly visible. The suspension is initially homogeneous and the trajectories of the particles are tracked after $\unit[10]{min}$ for about $\unit[30]{\second}$.}
\end{figure}

To demonstrate such particle separation, we have carried out additional experiments with a homogeneous suspension of self-propelled Janus particles with radius $R_1 = \unit[0.95]{\micro\metre}$ and a now $\unit[20]{nm}$ thick platinum cap. After filling a cuvette with the suspension and waiting for $\unit[10]{min}$, trajectories were recorded for colloids near the center of the $\unit[700]{\micro\metre}$ thick suspension and for colloids near the upper cuvette wall. Figure \ref{fig:Separation}a shows the probability distribution of the two-dimensional angular speed $\omega_{\mathrm{xy}}$ (see Sec.\ \ref{subsec:theory}) for the particles in the center of the suspension and near the upper interface. 
The colloids near the upper interface were found to have a small or no angular speed ($0\leqslant\omega_{\mathrm{xy}}\leqslant\unit[1]{s^{-1}}$), while those remaining in the bulk had a much wider distribution of the angular speed ($0\leqslant\omega_{\mathrm{xy}}\leqslant\unit[4]{s^{-1}}$). In a control experiment, we verified that interactions with the interface were not responsible for the reduction in $\omega_{\mathrm{xy}}$: eventually rapid rotors were observed near the cuvette interfaces and continued to show trajectories with loops in quick succession. Furthermore, the mean translational velocities for the bulk and interface colloids were similar ($v_{\mathrm{bulk}}=\unit[5.8]{\micro\metre/\second}$, $v_{\mathrm{interface}}=\unit[6.2]{\micro\metre/\second}$), suggesting that the spatial separation was due to the difference in $\omega_{0}$ rather than $v_{0}$.

To verify that mixtures of self-propulsive Janus spheres with varying angular self-propulsion undergo rapid stratification, we carried out additional simulations that correspond to the experiments.
Different from our previous simulations described in Sec.\ \ref{subsec:theory}, we chose the random initial particle positions with the $z$ coordinates uniformly distributed in the interval $[-\unit[350]{\micro\metre},\unit[350]{\micro\metre}]$ and simulated the trajectories for a $\unit[10]{min}$ longer period. 
After the initial $\unit[10]{min}$, where we discarded the particle coordinates, we tracked them for the remaining about $\unit[30]{\second}$, but only within two $\unit[100]{\micro\metre}$ thick horizontal layers in the center and at the upper interface of the suspension. 
In this way, we simulated the trajectories of $50,000$ particles with radius $R_1 = \unit[0.95]{\micro\metre}$, where we modeled the upper and lower interface of the suspension by hard horizontal plane walls.
Furthermore, in accordance with our experiments we used the translational self-propulsion speed $v_0=\unit[6]{\micro\metre/\second}$, sedimentation speed $v_{\mathrm{s}}=\unit[0.46]{\micro\metre/\second}$, angular self-propulsion speeds $\omega_{0}\in [0,\unit[5]{\second^{-1}}]$, and maximal aligning angular velocity $\omega_{\mathrm{M}}=\unit[0.27]{\second^{-1}}$, where $v_{\mathrm{s}}$ and $\omega_{\mathrm{M}}$ take the increased thickness and weight of the $\unit[20]{nm}$ thick platinum cap into account. 
Apart from that, we used the same parameters and procedure as for our previous simulations of self-propelled Janus spheres with radius $R_1 = \unit[0.95]{\micro\metre}$ (see Sec.\ \ref{subsec:theory}).
The results of our simulations of a mixture of Janus spheres with varying angular self-propulsion are shown in Fig.\ \ref{fig:Separation}b. 
For the particles tracked near the upper interface of the suspension the distribution of the two-dimensional angular speed $\omega_{\mathrm{xy}}$ is clearly shifted to smaller values of $\omega_{\mathrm{xy}}$ compared to the distribution for the particles tracked in the center of the suspension. 
This is similar to our corresponding experimental results. However, a quantitative comparison of our experimental and simulation results is not possible here, since the particular shapes of the curves in Fig.\ \ref{fig:Separation}b result from the probability distributions of the parameters $\omega_{\mathrm{A},1}$, $\omega_{\mathrm{A},2}$, and $\omega_{\mathrm{A},3}$ used in our simulations (see Sec.\ \ref{subsec:theory}). 
The real probability distributions for these parameters are certainly different resulting in the shapes of the curves in Fig.\ \ref{fig:Separation}a differing from those in Fig.\ \ref{fig:Separation}b, but it was not possible to measure these parameters for the particles used in our experiments. 

The possibility to sort self-propelled Janus particles with respect to their angular self-propulsion is relevant, e.g., for applications where one wants to separate Janus particles with a damaged or irregularly shaped metal cap from more symmetric Janus particles. 
Extracting particles from the top of a container would select for linear translators suitable for transport tasks, while the remaining particles would on average possess greater angular self-propulsion and so be suitable for mixing tasks.\cite{Balk2014} This procedure could also allow the variety of trajectories observed for self-propelled colloid agglomerates to be similarly separated.\cite{Ebbens2010a}

\section{\label{sec:conclusions}Conclusions}
Using experiments and theoretical considerations based on appropriate Langevin equations, we have studied the 3D trajectories of self-propulsive colloidal Janus spheres with bottom-heaviness under gravity. The self-propulsion of such particles often goes along with both a translational and an angular velocity that are constant in the particle frame. 
Gravity influences the motion of our particles by a force that is downwards oriented and by a torque that tends to align the particles upwards.

We found that these particles typically move along a helical path that is perturbed by Brownian noise. If the aligning torque resulting from the mass anisotropy outweighs the effect of rotational diffusion, the particle motion is gravitactic and the axis of the helical trajectory coincides with the direction of gravity. Otherwise, if the orienting torque is not dominant, no clear gravitational alignment is observed. As an additional finding, the characteristic features of the trajectory of a self-propelled Janus particle are largely governed by its angular self-propulsion velocity. If the latter exceeds a certain threshold value and is not parallel to the translational self-propulsion velocity, the swimming direction of the particle is perpetually rotated, leading to a trajectory with an endless series of loops rapidly following each other. As opposed to this, the trajectory of a particle with weak angular self-propulsion is more elongated or even straight, except for the influence of the Brownian noise.
The observed alignment mechanism under gravity and the strong dependence of the trajectory shape on the angular self-propulsion velocity are valuable features in the context of application purposes, since they allow to separate self-propelled colloidal Janus spheres according to their mass anisotropy and rotational activity. 

Our results are applicable not only to the specific Janus particles studied here but also to various other types of synthetic self-propelled particles. Since our theoretical considerations are not limited to chemically driven particles, the trajectories that we observed should also occur for Janus spheres that self-propel under illumination \cite{VolpeBVKB2011,Kroy2016,MoysesPSG2016,Lozano2016}. Furthermore, our findings can be generalized to other particle shapes. A first step would be to consider rotationally symmetric shape-anisotropic particles such as rods \cite{Paxton2004,Peruani2006} or spheroids \cite{Han2006,Passow2015,Kurzthaler2016}. For these systems, the description has to be adapted accordingly, but the underlying physics is basically the same. This is different for particles without any symmetry axis \cite{KuemmeltHWBEVLB2013,Chakrabarty2014}. In such situations, an intricate interplay between the gravitational effects resulting from an inhomogeneous mass distribution as reported in the present article and gravitactic behavior due to the shape asymmetry \cite{tenHagenKWTLB2014} will govern the particle dynamics.

Instead of exposing mass-anisotropic self-propelled particles to gravity, one could also consider such particles in a centrifuge. This allows to improve particle separation based on the reported separation phenomena by tuning the rotation speed of the centrifuge. 

The dynamic behavior of the self-propelled Janus particles described in this article is in close analogy to phenomena seen for swimming microorganisms.\cite{Jennings1901,Crenshaw1996}
With regard to the motion of microorganisms in nature, it would be an interesting task for the future to extend our work towards mass-anisotropic self-propelled colloidal particles in shear flow,\cite{tenHagenWL2011,Zoettl2012,TaramaMtHWOL2013} where we expect even more complex trajectories.

\begin{acknowledgments}
We thank Jonathan Howse for providing the 2D particle tracking algorithms and loan of a fluorescent filter. 
A.I.C.\ and S.E.\ thank the EPSRC for supporting this work via S.E.'s Career Acceleration Fellowship---EP/J002402/1. 
R.W., B.t.H., and H.L.\ are funded by the Deutsche Forschungsgemeinschaft (DFG, German Research Foundation)---WI 4170/3-1; HA 8020/1-1; LO 418/17-1.
\end{acknowledgments}

\appendix

\section{Determination of the sedimentation speed $\boldsymbol{v_{\mathrm{s}}}$}
Since our spherical Janus particles are self-propelled in $\mathrm{H}_{2}\mathrm{O}_{2}$, we cannot measure their sedimentation speed $v_{\mathrm{s}}$ directly when they are dispersed in the $\unit[10]{\%}$ mass fraction solution of $\mathrm{H}_{2}\mathrm{O}_{2}$. Therefore, we measured the particles' sedimentation speed in water $v_{\mathrm{s,W}}$ and calculated from $v_{\mathrm{s,W}}$ their sedimentation speed $v_{\mathrm{s}}$ in the $\mathrm{H}_{2}\mathrm{O}_{2}$ solution. 

The experimental results for $v_{\mathrm{s,W}}$ for particles with radii $R\in\{\unit[0.95]{\micro\metre},\unit[1.55]{\micro\metre},\unit[2.40]{\micro\metre}\}$ and a $d=\unit[10]{\nano\metre}$ thick platinum cap (see also Ref.\ [\onlinecite{Campbell2013}]) as well as with radius $R=\unit[0.95]{\micro\metre}$ and a $d=\unit[20]{\nano\metre}$ thick platinum cap are given in Table \ref{tab:Parameters}.
Note that $d$ is the maximal thickness of the platinum cap, which the cap attains at its pole. From there, the thickness is continuously tapering towards the edge of the hemispherical cap.\cite{Campbell2013}

Using the Stokes expression for the speed of a sphere settling in a viscous liquid, $v_{\mathrm{s,W}}$ can be expressed as 
\begin{equation}
v_{\mathrm{s,W}} = \frac{2}{9}\frac{\rho_{\mathrm{P}}-\rho_{\mathrm{W}}}{\eta_{\mathrm{W}}} g R^2 
\label{Stokes}%
\end{equation}
with the particle's average mass density $\rho_{\mathrm{P}}$, the mass density of water $\rho_{\mathrm{W}}=\unit[998]{\kilo\gram/\metre^{3}}$, the dynamic viscosity of water $\eta_{\mathrm{W}}=\unit[10^{-3}]{\pascal\,\second}$, and the gravitational acceleration $g=\unit[9.81]{\metre/\second^{2}}$.  
Equation \eqref{Stokes} allows to calculate $\rho_{\mathrm{P}}$ from $v_{\mathrm{s,W}}$ via
\begin{equation}
\rho_{\mathrm{P}}=\rho_{\mathrm{W}}+\frac{9\eta_{\mathrm{W}}}{2g R^2} v_{\mathrm{s,W}} \,.
\label{Density}%
\end{equation}
The resulting values for $\rho_{\mathrm{P}}$ are given in Table \ref{tab:Parameters}. 
Applying again the Stokes expression, we can now calculate $v_{\mathrm{s}}$ by 
\begin{equation}
v_{\mathrm{s}} = \frac{2}{9}\frac{\rho_{\mathrm{P}}-\rho_{\mathrm{L}}}{\eta} g R^2 = \frac{\eta_{\mathrm{W}}}{\eta} v_{\mathrm{s,W}} - \frac{2g R^2}{9\eta} (\rho_{\mathrm{L}}-\rho_{\mathrm{W}}) \,,
\label{vs}%
\end{equation}
where $\rho_{\mathrm{L}}=\unit[1038]{\kilo\gram/\metre^{3}}$ is the mass density of the $\mathrm{H}_{2}\mathrm{O}_{2}$ solution and $\eta=1.02\cdot\unit[10^{-3}]{\pascal\,\second}$ is its dynamic viscosity. This yields the values for $v_{\mathrm{s}}$ given in Table \ref{tab:Parameters}.     

\begin{table*}[tb]
\centering
\begin{ruledtabular}
\begin{tabular}{ccccccc}%
$R/\unit{\micro\metre}$ & $d/\unit{\nano\metre}$ & $l/R$ & $v_{\mathrm{s,W}}/(\unit{\micro\metre\,\second^{-1}})$ & $\rho_{\mathrm{P}}/(\unit{\kilo\gram\,\metre^{-3}})$ & $v_{\mathrm{s}}/(\unit{\micro\metre\,\second^{-1}})$ & $\omega_{\mathrm{M}}/\unit{\second^{-1}}$ \\
\hline
$0.95$ & $10$ & $0.75$ & $0.193$ & $1096$ & $0.1121$ & $0.0528$ \\
$1.55$ & $10$ & $0.75$ & $0.532$ & $1100$ & $0.3162$ & $0.0927$ \\
$2.40$ & $10$ & $0.75$ & $1.008$ & $1078$ & $0.4958$ & $0.0817$ \\
$0.95$ & $20$ & $0.76$ & $0.549$ & $1277$ & $0.4611$ & $0.2659$ \\
\end{tabular}
\end{ruledtabular}
\caption{\label{tab:Parameters}Parameters characterizing the self-propelled Janus particles with different radii $R$ and maximal thicknesses $d$ of the platinum cap: 
distance $l$ of the center of mass of the platinum cap from the geometric center of the particle, sedimentation speed $v_{\mathrm{s,W}}$ in water, average mass density $\rho_{\mathrm{P}}$, sedimentation speed $v_{\mathrm{s}}$ in the $\mathrm{H}_{2}\mathrm{O}_{2}$ solution, and maximal aligning angular velocity $\omega_{\mathrm{M}}$.}%
\end{table*}

\section{Calculation of the maximal aligning angular velocity $\boldsymbol{\omega_{\mathrm{M}}}$}
The anisotropic mass distribution of our Janus spheres leads to a gravitational torque that acts on the particles and tends to orient them upwards. Choosing the geometric center of a particle as reference point, the magnitude of this torque is given by
\begin{equation}%
M = l F \sin(\theta) \,,
\label{TorqueGeneral}%
\end{equation}%
where $l$ is the distance of the center of mass of the platinum cap from the geometric center of the particle, $F=(\rho_{\mathrm{Pt}}-\rho_{\mathrm{L}})V_{\mathrm{cap}} g$ is the modulus of the buoyancy-corrected gravitational force that acts on the cap, and $\theta\in[0,\pi]$ is the tilt angle defined in Fig.\ 1 of the main text.
Finally, $\rho_{\mathrm{Pt}}=\unit[21,450]{\kilo\gram/\metre^{3}}$ is the mass density of platinum and $V_{\mathrm{cap}}$ is the volume of the particle's platinum cap.

Some of us have previously shown that $l=0.75R$ for particles with a $d=\unit[10]{\nano\metre}$ thick platinum cap.\cite{Campbell2013} 
Using the cap model from Ref.\ [\onlinecite{Campbell2013}], we obtain $l\approx 0.76R$ for particles with $d=\unit[20]{\nano\metre}$ (see Table \ref{tab:Parameters}). 
To determine $V_{\mathrm{cap}}$, we use the expression 
\begin{equation}
\rho_{\mathrm{P}}=\frac{V_{\mathrm{S}}\rho_{\mathrm{PS}}+V_{\mathrm{cap}}\rho_{\mathrm{Pt}}}{V_{\mathrm{S}}+V_{\mathrm{cap}}} 
\label{rhoP}%
\end{equation}
with the volume $V_{\mathrm{S}}=4\pi R^{3}/3$ of the polystyrene sphere being part of the particle and the mass density $\rho_{\mathrm{PS}}=\unit[1050]{\kilo\gram/\metre^{3}}$ of polystyrene. 
This yields 
\begin{equation}
V_{\mathrm{cap}} = \frac{\rho_{\mathrm{P}}-\rho_{\mathrm{PS}}}{\rho_{\mathrm{Pt}}-\rho_{\mathrm{P}}} V_{\mathrm{S}} \,.
\label{Vcap}%
\end{equation}

For $\theta=\pi/2$, the torque $M$ reaches its maximal value $M_{\mathrm{max}}=lF$.
The maximal aligning angular velocity $\omega_{\mathrm{M}}$, which corresponds to $M_{\mathrm{max}}$, is given by 
\begin{equation}%
\omega_{\mathrm{M}} = \frac{D_{\mathrm{R}}M_{\mathrm{max}}}{k_{\mathrm{B}}T} = l \frac{\rho_{\mathrm{Pt}}-\rho_{\mathrm{L}}}{8\pi\eta R^{3}} g V_{\mathrm{cap}}
\label{omegaMI}%
\end{equation}%
with the particle's rotational diffusion coefficient $D_{\mathrm{R}}=k_{\mathrm{B}}T/(8\pi\eta R^{3})$, Boltzmann constant $k_{\mathrm{B}}$, and temperature $T$.
Inserting Eq.\ \eqref{Vcap} into Eq.\ \eqref{omegaMI} leads to 
\begin{equation}%
\omega_{\mathrm{M}} = \frac{lg}{6\eta} \frac{\rho_{\mathrm{Pt}}-\rho_{\mathrm{L}}}{\rho_{\mathrm{Pt}}-\rho_{\mathrm{P}}} (\rho_{\mathrm{P}}-\rho_{\mathrm{PS}}) \,.
\label{omegaMII}%
\end{equation}%
Together with Eq.\ \eqref{Density} one obtains from Eq.\ \eqref{omegaMII} the values for $\omega_{\mathrm{M}}$ given in Table \ref{tab:Parameters}.

\bibliographystyle{apsrev4-1}
\bibliography{bibliography}
\end{document}